\documentclass[%
 reprint,
 amsmath,amssymb,
 aps,
]{revtex4-2}

\usepackage{graphicx}
\usepackage{dcolumn}
\usepackage{bm}
\usepackage{xcolor}

\begin{document}

\preprint{APS/123-QED}

\title{\textcolor{black}{Quantifying Energy Conversion in \\ Higher Order Phase Space Density Moments in Plasmas}}

\author{Paul A.~Cassak}
\email{Paul.Cassak@mail.wvu.edu}
\author{M.~Hasan Barbhuiya}
\affiliation{Department of Physics and Astronomy and the Center for KINETIC Plasma Physics \\West Virginia University, Morgantown, WV 26506, USA}

\author{Haoming Liang}
\affiliation{Center for Space Plasma and Aeronomic Research \\University of Alabama in Huntsville, Huntsville, AL 35899, USA}

\author{Matthew R.~Argall}
\affiliation{Space Science Center, Institute for the Study of Earth, Oceans, and Space \\ University of New Hampshire, Durham, NH 03824, USA}

\date{\today}

\begin{abstract}
Weakly collisional and collisionless plasmas are typically far from local thermodynamic equilibrium (LTE), and understanding energy conversion in such systems is a forefront research problem.  The standard approach is to investigate changes in internal (thermal) energy and density, but this omits energy conversion that changes any higher order moments of the phase space density. In this study, we calculate from first principles the energy conversion associated with all higher moments of the phase space density for systems not in LTE.  Particle-in-cell simulations of collisionless magnetic reconnection reveal that energy conversion associated with higher order moments can be locally significant. The results may be useful in numerous plasma settings, such as reconnection, turbulence, shocks, and wave-particle interactions in heliospheric, planetary, and astrophysical plasmas.
\end{abstract}

\maketitle


  
Energy conversion is largely well understood for systems with initial and final states in or near local thermodynamic equilibrium (LTE) \cite{Chapman70,Jou10}. However, energy conversion in systems far from LTE, such as weakly collisional or collisionless plasmas endemic to many space and astrophysical environments, remains a forefront research area \cite{Howes17,Matthaeus20}.

For a species $\sigma$ not in LTE, internal moments of the phase space density $f_\sigma$ are defined as $f_\sigma$ multiplied by powers of components of ${\bf v}_\sigma^\prime$ and integrated over all velocity space. Here, the random velocity is ${\bf v}_\sigma^\prime = {\bf v} - {\bf u}_\sigma$, velocity space coordinate is ${\bf v}$, bulk flow velocity is ${\bf u}_\sigma = (1/n_\sigma) \int f_\sigma {\bf v} d^3v$, and number density is $n_\sigma = \int f_\sigma d^3v$. A standard approach to study energy conversion in plasmas \cite{Yang17,Chasapis18,Zhong19,Bandyopadhyay20,bandyopadhyay_energy_2021,Wang21,zhou_measurements_2021,del_sarto_pressure_2016,Sitnov18,del_sarto_pressure_2018,Du18,Parashar18,Pezzi19,yang_scale_2019,song_forcebalance_2020,Du20,Fadanelli21,Arro21,Yang_2022_ApJ,Hellinger22,Cassak_PiD1_2022,Cassak_PiD2_2022,Barbhuiya_PiD3_2022} centers on the first few internal moments. Compressional work describes changes to $n_\sigma$, {\it i.e.,} the zeroth internal moment of $f_\sigma$, described by the continuity equation \cite{Yang17,Braginskii65}. The internal energy per particle ${\cal E}_{\sigma,{\rm int}} = (3/2)k_B {\cal T}_\sigma$, {\it i.e.,} the second internal moment of $f_\sigma$ divided by $n_\sigma$, can change due to compressional heating by work $-{\cal P}_\sigma (\nabla \cdot {\bf u}_\sigma)$, incompressional heating via the remainder of the pressure-strain interaction (called Pi-D \cite{Yang17}), heat flux, or collisions, according to \cite{Braginskii65,Jou10,Yang17}
\begin{equation}
\frac{3}{2}n_\sigma k_B\frac{d{\cal T}_\sigma}{dt} = - ({\bf P}_\sigma \cdot \nabla) \cdot {\bf u}_\sigma - \nabla \cdot {\bf q}_\sigma + n_\sigma \dot{Q}_{\sigma,{\rm coll,inter}}. \label{eq:intenergyevolve}
\end{equation}
Here, the elements of the pressure tensor ${\bf P}_\sigma$ are $P_{\sigma,jk} = m_\sigma \int v_{\sigma j}^\prime v_{\sigma k}^\prime f_\sigma d^3v$, temperature tensor is ${\bf T}_\sigma = {\bf P}_\sigma / n_\sigma k_B$, 
effective pressure is ${\cal P}_\sigma = (1/3) {\rm tr}[{\bf P}_\sigma]$, effective temperature is ${\cal T}_\sigma = {\cal P}_\sigma / n_\sigma k_B = (m_\sigma / 3 n_\sigma k_B) \int v_\sigma^{\prime 2} f_\sigma d^3v$, vector heat flux density is ${\bf q}_\sigma = \int (1/2) m_\sigma v_\sigma^{\prime 2} {\bf v}_\sigma^\prime f_\sigma d^3v$, and volumetric heating rate per particle due to inter-species collisions is $\dot{Q}_{\sigma,{\rm coll,inter}} = (1/n_\sigma) \int (1/2) m_\sigma v_\sigma^{\prime 2} \sum_{\sigma^\prime} C_{{\rm inter}}[f_\sigma, f_{\sigma^\prime}] d^3v$, where the inter-species collision operator is $C_{{\rm inter}}[f_\sigma, f_{\sigma^\prime}]$, $k_B$ is Boltzmann's constant, $m_\sigma$ is the constituent mass, and $d/dt = \partial/\partial t + {\bf u}_\sigma \cdot \nabla$ is the convective derivative.
 
There is an energy conversion channel beyond those discussed thus far. $f_\sigma$ has an infinite number of internal moments that are all treated on equal footing. While Eq.~(\ref{eq:intenergyevolve}) includes the impact of off-diagonal pressure tensor elements and heat flux on ${\cal E}_{\sigma,{\rm int}}$, any energy conversion associated with time evolution of all other internal moments themselves is not contained in the continuity equation or Eq.~(\ref{eq:intenergyevolve}).

Studies have addressed time evolution of other moments and their contribution to energy conversion. The evolution of non-isotropic pressures has been studied \cite{Kuznetsova98,Yin03,Brackbill11,Greco12,Servidio12,Egedal13,Wang15,Swisdak16,del_sarto_pressure_2016,del_sarto_pressure_2018}. Other approaches capture the effect of all moments of $f_\sigma$. Linearizing $f_\sigma$ around its equilibrium in kinetic theory and gyrokinetics reveals the so-called free energy \cite{Hallatschek04,Howes06,Schekochihin09}, which quantifies non-LTE energy conversion into mechanical or magnetic energy \cite{Hallatschek04}. It is associated with the phase space cascade of entropy which can lead to dissipation \cite{Tatsuno09}. The velocity space cascade has been studied without linearizing $f_\sigma$ \cite{Servidio17,Pezzi18,Cerri18,Pezzi19}. In another approach, changes to bulk kinetic energy are quantified kinetically using field-particle correlations \cite{Klein16,Howes17b,Klein17,Klein17b,Chen19,Li2019collisionless,Klein20,Juno21,Verniero2021patch,Montag22}.

In this study, we use a first-principles theory to quantify energy conversion associated with all internal moments. We show this energy conversion is physically associated with changing the velocity space shape of $f_\sigma$. There are three important ingredients. First, the key quantity is kinetic entropy \cite{Boltzmann77,Eu95,Eu06,Jou10,Eyink18} rather than energy. Second, we employ the decomposition of kinetic entropy into position and velocity space kinetic entropy \cite{Mouhot11,Liang19}. Third, we employ the so-called relative entropy \cite{Grad65,Eu95,Eu06}. Our analysis was performed independently, but we found it is similar to treatments in chemical physics of dilute gases \cite{Eu95} and quantum statistical mechanics \cite{Floerchinger20}. The novelty of our analysis stems from using the decomposition of kinetic entropy and significant differences in interpretation than in previous work. We employ a particle-in-cell (PIC) simulation of collisionless magnetic reconnection, revealing energy conversion associated with higher order moments can be locally significant.


We first derive an expression for the rate of energy conversion associated with non-LTE internal moments of $f_\sigma$, emphasizing departures from the treatment in Ref.~\cite{Eu95}.  We assume a classical (non-relativistic, non-quantum) three-dimensional (3D) system of infinite volume or in a thermally insulated domain with a fixed number $N_\sigma$ of monatomic particles.  The kinetic entropy density $s_\sigma$ associated with $f_\sigma$ is \cite{Boltzmann1872}
\begin{equation}
    s_\sigma = - k_{B} \int f_\sigma \ln\left(\frac{f_\sigma \Delta^3r_\sigma \Delta^3v_\sigma}{N_\sigma}\right) d^3v, \label{eq:entropydensitydef}
\end{equation}
where the integral is over all velocity space, and $\Delta^3r_\sigma$ and $\Delta^3v_\sigma$ are position space and velocity space volume elements in phase space, respectively \cite{Liang19,Liang20,Argall22}.  In the comoving (Lagrangian) frame, $s_\sigma$ evolves according to (\cite{Eu95} and Supplemental Material A \cite{Cassak2022_EntropyPRL_Supp})
\begin{equation}
    \frac{d}{dt} \left(\frac{s_\sigma}{n_\sigma}\right) + \frac{\nabla \cdot \bm{\mathcal{J}}_{\sigma,{\rm th}}}{n_\sigma} = \frac{\dot{s}_{\sigma,{\rm coll}}}{n_\sigma}, \label{eq:entdensev3}
\end{equation}
where $\bm{\mathcal{J}}_{\sigma,{\rm th}}$ is thermal kinetic entropy density flux and $\dot{s}_{\sigma,{\rm coll}}$ is local time rate of change of kinetic entropy density through collisions, defined in Eqs.~(S.4) and (S.3), respectively.  We note that Eq.~(\ref{eq:entdensev3}) has no explicit dependence on body forces including gravitational and electromagnetic forces, which implies they do not directly change internal moments of $f_\sigma$.  Eq.~(\ref{eq:intenergyevolve}) exemplifies this for the special case of internal energy.

In a key departure from Ref.~\cite{Eu95}, we decompose kinetic entropy density $s_\sigma$ into a position space kinetic entropy density $s_{\sigma p}$ and velocity space kinetic entropy density $s_{\sigma v}$, with $s_\sigma = s_{\sigma p} + s_{\sigma v}$, as \cite{Mouhot11,Liang19}
\begin{subequations}
\begin{eqnarray}
    s_{\sigma p} &=& -k_{B} n_\sigma \ln \left(\frac{n_\sigma \Delta^3r_\sigma}{N_\sigma}\right), \label{eq:sposition} \\ s_{\sigma v} &=& -k_{B} \int f_\sigma \ln \left(\frac{f_\sigma \Delta^{3}v_\sigma}{n_\sigma}\right)d^{3}v. \label{eq:svelocity}
\end{eqnarray}
\end{subequations} 
A direct calculation (see Supplemental Material B-D) of the terms on the left side of Eq.~(\ref{eq:entdensev3}) using Eqs.~(\ref{eq:sposition}) and (\ref{eq:svelocity}) gives
\begin{subequations}
\begin{eqnarray}
      \frac{d}{dt} \left( \frac{s_{\sigma p}}{n_\sigma} \right) & = & \frac{1}{{\cal T}_\sigma} \frac{d{\cal W}_\sigma}{dt}, 
      \label{eq:ddtspovern} \\ 
\frac{d}{dt} \left( \frac{s_{\sigma v}}{n_\sigma} \right) & = & \frac{1}{{\cal T}_\sigma} \frac{d {\cal E}_{\sigma,{\rm int}}}{dt} + \frac{d}{dt} \left( \frac{s_{\sigma v,{\rm rel}}}{n_\sigma} \right), \label{eq:ddtsceq} \\
\frac{\nabla \cdot \bm{\mathcal{J}}_{\sigma,{\rm th}}}{n_\sigma} & = & - \frac{1}{{\cal T}_\sigma} \frac{d{\cal Q}_\sigma}{dt} + \frac{(\nabla \cdot \bm{\mathcal{J}}_{\sigma,{\rm th}})_{{\rm rel}}}{n_\sigma},\label{eq:divjthon}
\end{eqnarray}
\end{subequations}
where $d{\cal W}_\sigma = {\cal P}_\sigma d(1/n_\sigma)$ is the compressional work per particle done by the system, $d{\cal E}_{\sigma,{\rm int}} = (3/2) k_B d{\cal T}_\sigma$ is the increment in internal energy per particle, and $d{\cal Q}_\sigma/dt = [-\nabla \cdot {\bf q}_\sigma - ({\bf P}_{\sigma} \cdot \nabla) \cdot {\bf u}_\sigma + {\cal P}_\sigma (\nabla \cdot {\bf u}_\sigma)]/n_\sigma$ is the (thermodynamic) heating rate per particle from sources other than compression that can change the effective temperature [see Eq.~(\ref{eq:intenergyevolve})]. Lastly, $s_{\sigma v,{\rm rel}}$ is the relative entropy density  and $(\nabla \cdot \bm{\mathcal{J}}_{\sigma,{\rm th}})_{{\rm rel}}$ is the thermal relative entropy density flux divergence, given by
\begin{eqnarray}
s_{\sigma v,{\rm rel}} & = & -k_B \int f_\sigma \ln \left(\frac{f_\sigma}{f_{\sigma M}}\right) d^3v, \label{eq:svrel} \\
(\nabla \cdot \bm{\mathcal{J}}_{\sigma,{\rm th}})_{{\rm rel}} & = & -k_{B} \int \left[ \frac{}{} \nabla \cdot ({\bf v}_\sigma^\prime f_\sigma)\right] \ln\left(\frac{f_\sigma}{f_{\sigma M}}\right) d^{3} v, \label{eq:divjthrel}
\end{eqnarray}
and the ``Maxwellianized'' phase space density $f_{\sigma M}$ associated with $f_\sigma$ is \cite{Grad65}
\begin{equation}
  f_{\sigma M} = n_\sigma \left(\frac{m_\sigma}{2\pi k_B {\cal T}_\sigma}\right)^{3/2} e^{-m_\sigma({\bf v} - {\bf u}_\sigma)^2/2 k_B {\cal T}_\sigma}, \label{eq:maxwellian}
\end{equation}
where $n_\sigma$, ${\bf u}_\sigma$, and ${\cal T}_\sigma$ are based on $f_\sigma$. (Ref.~\cite{Eu95} used a more general reference phase space density than $f_{\sigma M}$, so our choice is a special case of theirs.) 

Equations~(\ref{eq:ddtspovern})-(\ref{eq:divjthon}) have important implications, and our interpretation greatly departs from Ref.~\cite{Eu95}. Ignoring the relative terms in Eqs.~(\ref{eq:ddtsceq}) and (\ref{eq:divjthon}), we see Eq.~(\ref{eq:entdensev3}) (scaled by the effective temperature) inherently contains information about work, internal energy, and thermodynamic heat as captured by the continuity equation and Eq.~(\ref{eq:intenergyevolve}).  This suggests the relative terms describe energy conversion associated with all internal moments beyond the second moment.

We therefore define increments of relative energy per particle $d{\cal E}_{\sigma,{\rm rel}}$ and relative heat per particle $d{\cal Q}_{\sigma,{\rm rel}}$ by
\begin{subequations}
\begin{eqnarray}
    \frac{d{\cal E}_{\sigma,{\rm rel}}}{dt} & = & {\cal T}_\sigma \frac{d(s_{\sigma v,{\rm rel}}/n_\sigma)}{dt}, \label{eq:msigreldef} \\
    \frac{d{\cal Q}_{\sigma,{\rm rel}}}{dt} & = & -{\cal T}_\sigma \frac{(\nabla \cdot \bm{\mathcal{J}}_{\sigma,{\rm th}})_{{\rm rel}}}{n_\sigma}. \label{eq:ddtqreldef}
\end{eqnarray}
\end{subequations}
Further defining energy increments per particle in all internal moments at and above the second moment as $d{\cal E}_{\sigma,{\rm gen}} = d{\cal E}_{\sigma,{\rm int}} + d{\cal E}_{\sigma,{\rm rel}}$ and generalized heat per particle as $d{\cal Q}_{\sigma,{\rm gen}} = d{\cal Q}_{\sigma} + d{\cal Q}_{\sigma,{\rm rel}}$, Eqs.~(\ref{eq:entdensev3}) - (\ref{eq:divjthon}), (\ref{eq:msigreldef}) and (\ref{eq:ddtqreldef}) take on the simple form
\begin{equation}
\frac{d{\cal W}_\sigma}{dt} + \frac{d{\cal E}_{\sigma,{\rm gen}}}{dt} = \frac{d{\cal Q}_{\sigma,{\rm gen}}}{dt} + \dot{{\cal Q}}_{\sigma,\rm coll}. \label{eq:firstlawfinal}
\end{equation}
Equation~(\ref{eq:firstlawfinal}) generalizes Eq.~(\ref{eq:intenergyevolve}), which contains energy conversion associated with only density and effective temperature, as opposed to all internal moments of $f_\sigma$. This interpretation is a significant departure from Ref.~\cite{Eu95}.

\begin{figure}
    \begin{center}\includegraphics[width=3.4in]{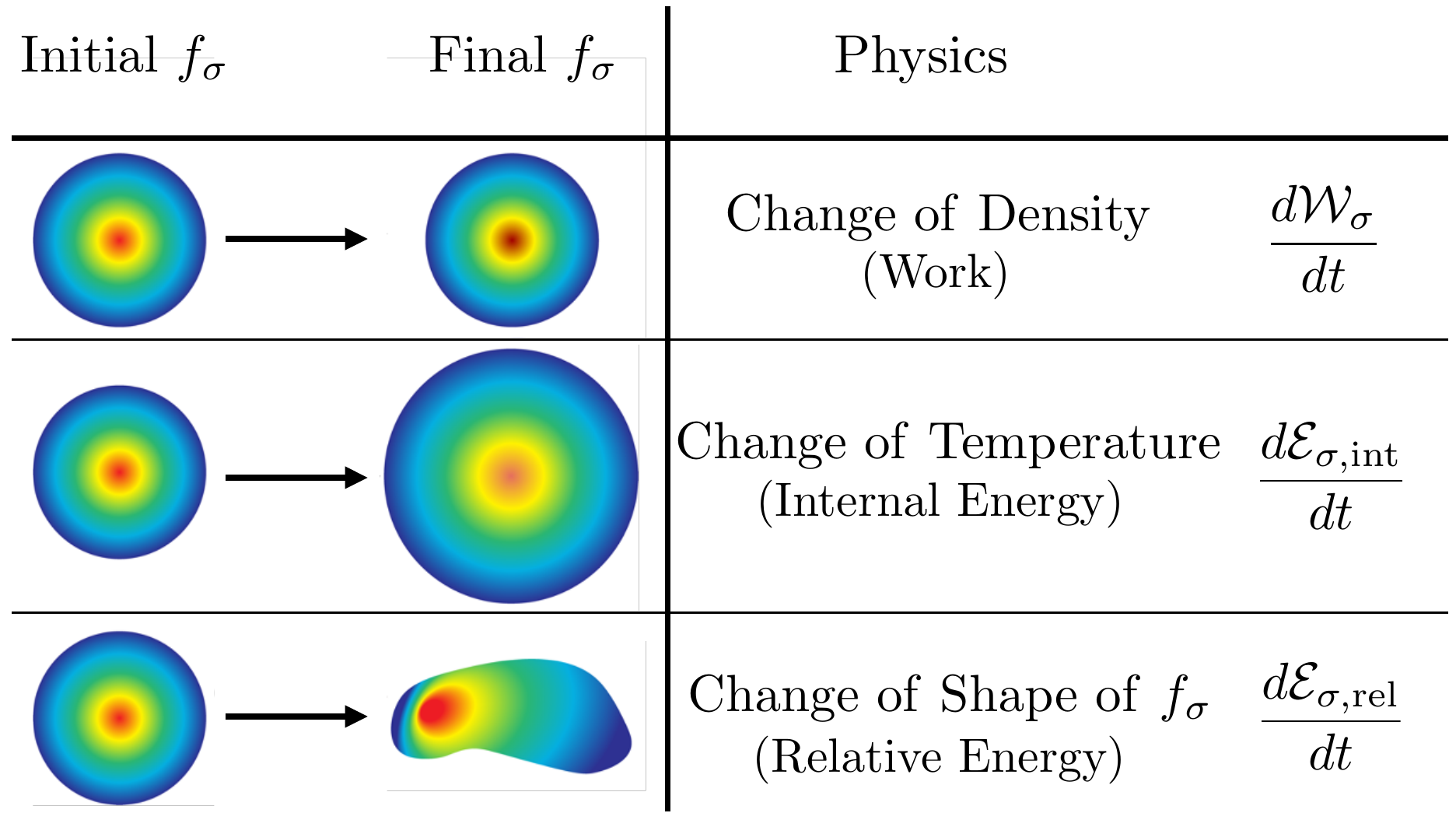}
    \caption{Schematic showing energy conversion channels according to their impact on the phase space density $f_\sigma$. The initial $f_\sigma$ is depicted as Maxwellian for illustrative purposes on the left. The final $f_\sigma$ is to their right. The descriptions of the changes in $f_\sigma$ are to their right.}
    \label{fig:distfuncshape}
    \end{center}
\end{figure}

We now provide a physical interpretation, which requires understanding energy conversion via its impact on $f_\sigma$. Work per particle $d{\cal W}_\sigma = {\cal P}_\sigma d(1/n_\sigma)$ changes the zeroth moment of $f_\sigma$. This is depicted graphically in Fig.~\ref{fig:distfuncshape}, where two velocity space dimensions of $f_\sigma$ are sketched. The top row shows a process taking a Maxwellianized $f_\sigma$ from an initial to final state. The intensification of colors denote a change in $f_\sigma$, and therefore $n_\sigma$. Similarly, $d{\cal E}_{\sigma,{\rm int}}$ is associated with changes to the second internal moment of $f_\sigma$, depicted in the second row of Fig.~\ref{fig:distfuncshape} for a process that increases ${\cal E}_{\sigma,{\rm int}}$, {\it i.e.,} the Maxwellianized $f_\sigma$ spreads in velocity space. 

To interpret $d{\cal E}_{\sigma,{\rm rel}}$, Eq.~(\ref{eq:svrel}) shows $s_{\sigma v,{\rm rel}}$ vanishes if $f_\sigma$ is a Maxwellian ($f_\sigma = f_{\sigma M})$ \cite{Grad65}. Thus, $d{\cal E}_{\sigma,{\rm rel}}$ describes non-LTE physics. Since a Maxwellian is the highest kinetic entropy state for a fixed $N_\sigma$ and ${\cal E}_{\sigma,{\rm int}}$~\cite{Boltzmann77}, $d(s_{\sigma v,{\rm rel}}/n_\sigma)/dt > 0$ implies $f_\sigma$ evolves towards Maxwellianity in the comoving frame, associated with $d{\cal E}_{\sigma,{\rm rel}} > 0$, while $d(s_{\sigma v,{\rm rel}}/n_\sigma)/dt < 0$ implies $f_\sigma$ evolves away from Maxwellianity and $d{\cal E}_{\sigma,{\rm rel}} < 0$. A process changing the shape of $f_\sigma$ is depicted in the third row of Fig.~\ref{fig:distfuncshape}, where $f_\sigma$ is initially Maxwellian and finally it is not.  

A concrete example showing that $d{\cal E}_{\sigma,{\rm rel}}$ is associated with $f_\sigma$ changing shape is provided in Supplemental Material E. $d{\cal E}_{\sigma,{\rm rel}}$ is calculated analytically for a bi-Maxwellian distribution with converging flow. It is shown that the evolution of $f_\sigma$ is consistent with the interpretation in the previous paragraph.

Collisions directly change the shape of $f_\sigma$, so $d{\cal E}_{\sigma,{\rm rel}}$ includes irreversible contributions if collisions are present. However, since $f_\sigma$ can change shape even in the perfectly collisionless limit, $d{\cal E}_{\sigma,{\rm rel}}$ also contains reversible effects.  Thus, the term is not purely irreversible as previously suggested \cite{Eu95}.

$d{\cal Q}_{\sigma}$ describes non-Maxwellian features of $f_\sigma$ that cause a flux of energy per particle that changes ${\cal T}_\sigma$ [see Eq.~(\ref{eq:intenergyevolve})]. $d{\cal Q}_{\sigma,{\rm rel}}$ is analogous: non-Maxwellian features in higher order internal moments produce a flux that modifies internal moments of $f_\sigma$ other than $n_\sigma$ and ${\cal T}_\sigma$. $\dot{Q}_{\sigma,{\rm coll}}$ describes both intra- and inter-species collisions, as opposed to solely inter-species arising in Eq.~(\ref{eq:intenergyevolve}). This is because both collision types can change higher order internal moments of $f_\sigma$, while elastic intra-species collisions conserve energy. 

\begin{figure*}
\begin{center}\includegraphics[width=\textwidth]{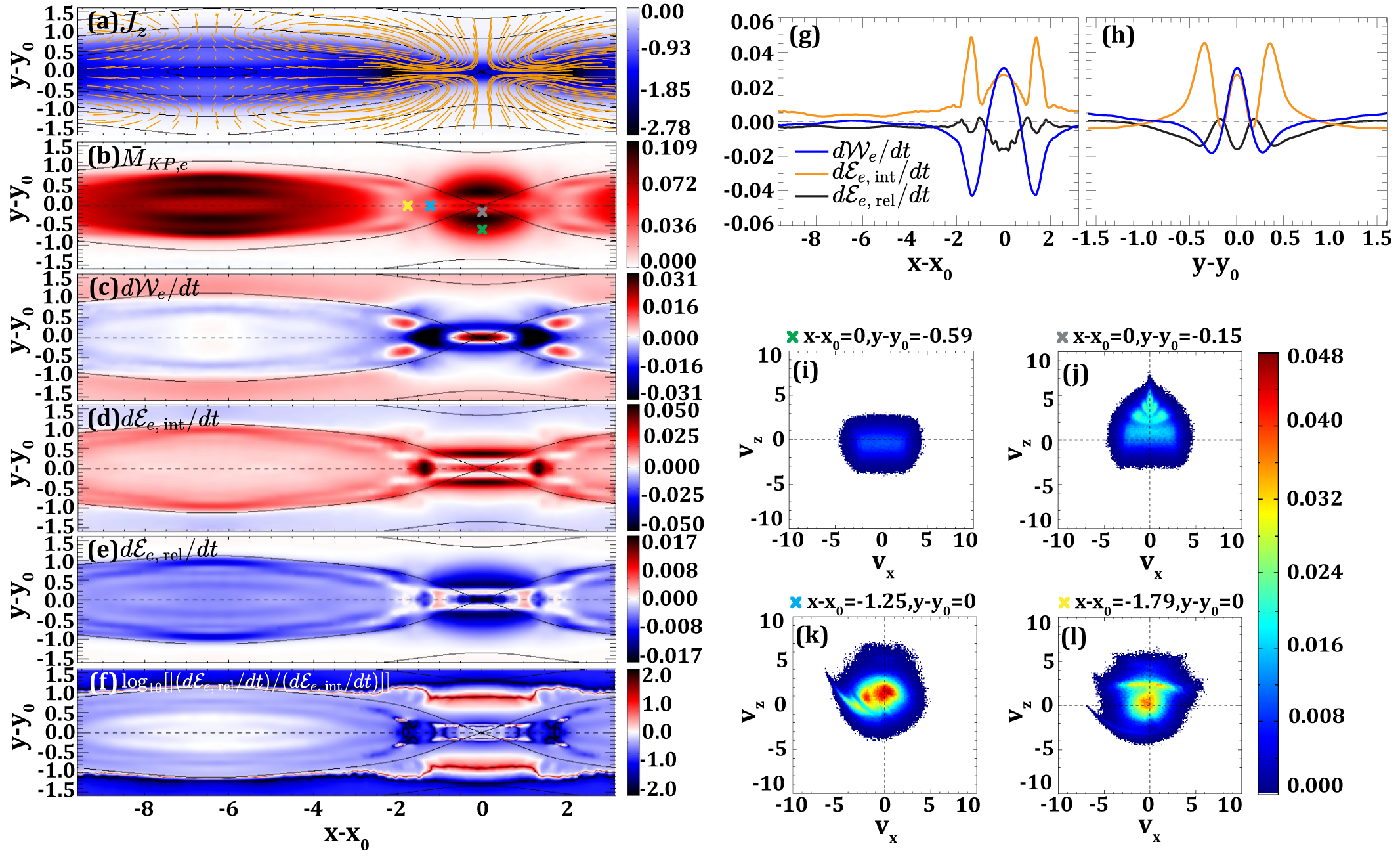}
\caption{Electron energy conversion in a PIC simulation of magnetic reconnection. (a) Out-of-plane current density $J_z$, with projections of magnetic field lines and segments of electron velocity streamlines overplotted in black and orange, respectively. (b) Electron entropy-based non-Maxwellianity $\bar{M}_{KP,e}$. Time rates of change per particle of (c) work $d{\cal W}_e/dt$, (d) internal energy $d{\cal E}_{e,{\rm int}}/dt$, and (e) relative energy $d{\cal E}_{e,{\rm rel}}/dt$. (f) $\log_{10}[|(d{\cal E}_{e,{\rm rel}}/dt)/(d{\cal E}_{e,{\rm int}}/dt)|]$. 1D cuts of the terms in panels (c)-(e) in the (g) $x$ and (h) $y$ directions. (i)-(l) Reduced electron phase space density $f_e(v_x,v_z)$ at locations denoted by the colored x's at the top left of the plots corresponding to the x's in panel (b) along a streamline.}
\label{fig:entrpyFluxQuantities}
\end{center}
\end{figure*}

We demonstrate key results of the theory using simulations of reconnection. Data are from the simulation in Ref.~\cite{Barbhuiya_PiD3_2022}. The code and numerical aspects are discussed there and in Supplemental Material F. The out-of-plane current density $J_z$ around a reconnection X-line at $(x_0,y_0)$ is in Fig.~\ref{fig:entrpyFluxQuantities}(a), with reversing magnetic field lines in black and electron streamline segments in orange, revealing typical profiles. 

We first confirm relative energy changes are related to $f_\sigma$ evolving towards or away from LTE. Figure~\ref{fig:entrpyFluxQuantities}(b) shows the electron entropy-based Kaufmann and Paterson non-Maxwellianity $\bar{M}_{e,KP} = (s_{e M}-s_e)/[(3/2) k_B n_e]$ \cite{Kaufmann09,Liang20}, where $s_e$ comes from Eq.~(\ref{eq:entropydensitydef}) based on $f_e$, while $s_{e M}$ comes from Eq.~(\ref{eq:entropydensitydef}) based on $f_{e M}$ in Eq.~(\ref{eq:maxwellian}).  It is a measure of the temporally and spatially local departure from LTE. Figure~\ref{fig:entrpyFluxQuantities}(e) is the rate of relative energy per particle $d{\cal E}_{e,{\rm rel}}/dt$. Figure~\ref{fig:entrpyFluxQuantities}(i)-(l) are reduced electron phase space densities $f_e(v_x,v_z)$ at the four color-coded x's along a streamline in Fig.~\ref{fig:entrpyFluxQuantities}(b).

$\bar{M}_{e,KP}$ and $d{\cal E}_{e,{\rm rel}}/dt$ together reveal whether $f_\sigma$ is locally in LTE [panel (b)] and whether it is evolving towards or away from LTE [(e)]. Just upstream of the electron diffusion region (EDR) ($|x-x_0| < 1, 0.45 < |y-y_0| < 1$), electrons get trapped by the upstream magnetic field \cite{Egedal13}, so $f_e$ becomes non-Maxwellian [dark red in (b)], with $f_e$ elongated in the parallel direction [(i)].  Thus, in the comoving frame, as a fluid element convects towards the X-line from upstream, $f_e$ evolves away from Maxwellianity, consistent with (e) where $d{\cal E}_{e,{\rm rel}}/dt < 0$.  Continuing towards the X-line, $f_e$ develops striations [(j)] due to electrons becoming demagnetized in the reversed magnetic field \cite{Speiser65,Ng11}.  This is associated with evolution away from LTE [blue in (e)]. Downstream of the X-line, there is a red patch in (e) at $|x-x_0| \simeq 1.25, |y-y_0| \simeq 0$ where electrons thermalize (Maxwellianize) \cite{Shuster14,Wang16}, which is seen in $f_e$ [(k)]. Just downstream from there ($|x-x_0| \simeq 1.8$), $f_e$ evolves away from LTE where electrons begin to remagnetize at the downstream edge of the EDR \cite{Shuster14,BarbhuiyaRing22} [(l)]. These results confirm the sign of $d{\cal E}_{\sigma,{\rm rel}}$ identifies whether $f_\sigma$ changes shape towards or away from LTE in the comoving frame.

Next, we demonstrate the quantitative importance of relative energy. Rates of work and internal energy per particle are shown in Figs.~\ref{fig:entrpyFluxQuantities}(c) and (d), respectively. Cuts of these quantities through the X-line in the horizontal and vertical directions, along with $d{\cal E}_{e,{\rm rel}}/dt$, are plotted in Figs.~\ref{fig:entrpyFluxQuantities}(g) and (h), respectively. At the X-line, the values are 0.031, 0.027, and $-0.016$, respectively, in normalized code units. Their sum, 0.042, is the total rate of energy per particle going into internal moments of electrons.  To see that relative energy is important, the standard approach using Eq.~(\ref{eq:intenergyevolve}) would say the energy rate going into changing $n_e$ and ${\cal T}_e$ is $0.031 + 0.027 = 0.058$, 38\% higher than the total rate when relative energy is included, which is a significant difference.

To assess its importance in other locations, Fig.~\ref{fig:entrpyFluxQuantities}(f) shows $\log_{10}[|(d{\cal E}_{e,{\rm rel}}/dt)/(d{\cal E}_{e,{\rm int}}/dt)|]$, with a color bar saturated at $\pm 2$ to better reveal details. Where internal and relative energy changes are comparable are white. Locations where $|d{\cal E}_{e,{\rm rel}}|$ exceeds $|d{\cal E}_{e,{\rm int}}|$ are red, especially just upstream of the EDR. In the deep blue regions, $|d{\cal E}_{e,{\rm rel}}| \ll |d{\cal E}_{e,{\rm int}}|$. In the light blue regions, including much of the EDR and island, $|d{\cal E}_{e,{\rm rel}}|$ is at least 20\% of the magnitude of $|d{\cal E}_{e,{\rm int}}|$. Thus, energy conversion associated with non-LTE internal moments in reconnection is broadly non-negligible, and can be locally significant or even dominant.


\begin{figure}
\begin{center}\includegraphics[width=3.4in]{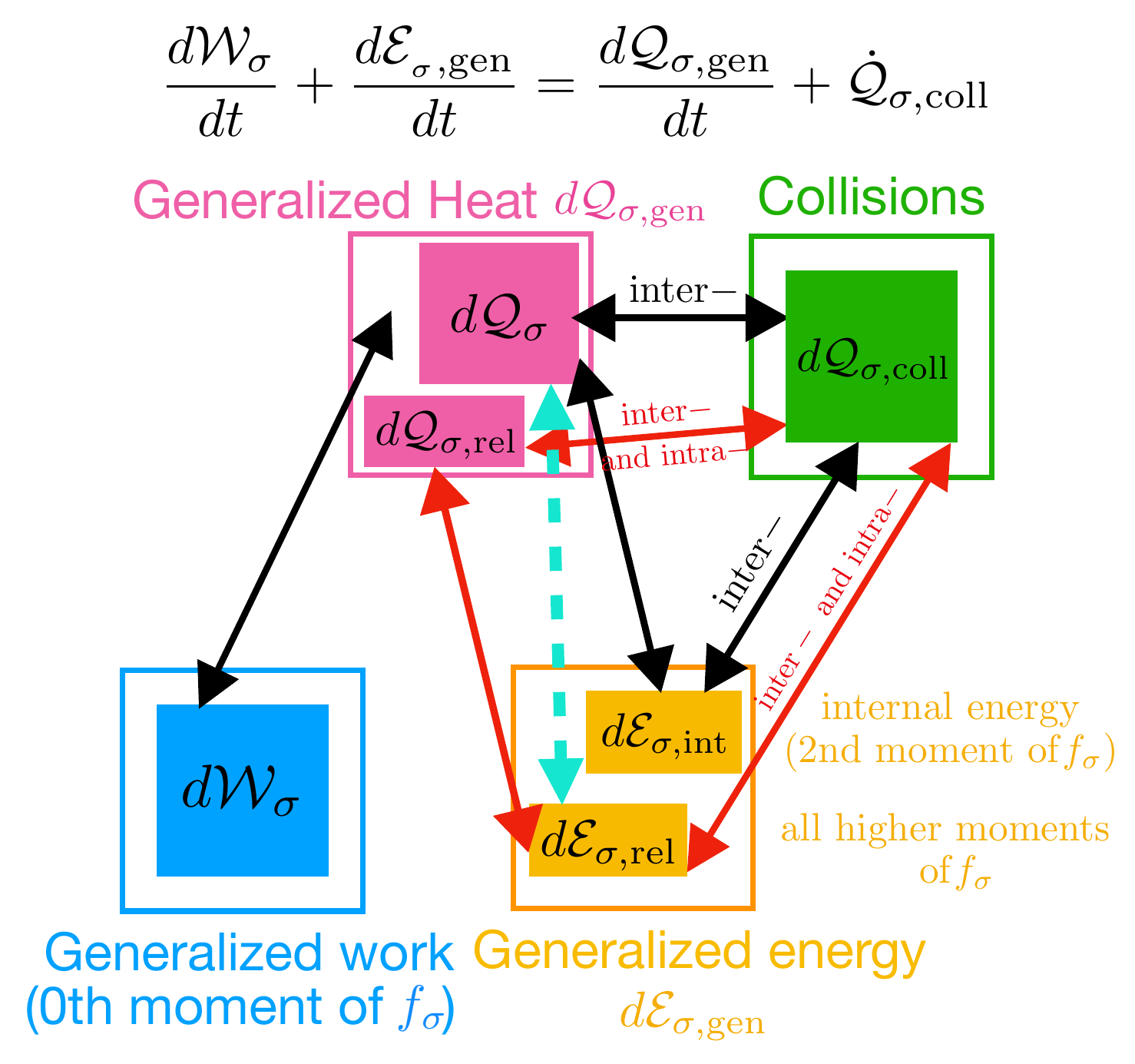}
\caption{Sketch illustrating energy conversion from Eq.~(\ref{eq:firstlawfinal}). Arrows show conversion channels between work (blue), heat (pink), energy (orange), and collisions (green), with standard channels in black and relative channels in red.  The light blue dashed arrow signifies how the relative terms couple to thermodynamic terms.}
\label{fig:firstlaw}
\end{center}
\end{figure}

We conclude with implications of the present results. First, the theory applies for systems arbitrarily far from LTE, so it could lead to significant advances compared to manifestly perturbative theories \cite{Jou10,Chapman70,Schekochihin09}. An extensive comparison to previous work is in Supplemental Material G. For a physical process that changes both internal energy and higher order moments, the theory captures both and allows each to be calculated separately. Since the theory contains all internal moments of $f_\sigma$, it overcomes the closure problem.

It is important to note that internal energy per particle ${\cal E}_{\sigma,{\rm int}}$ is a state variable, meaning it is history independent, but relative energy per particle ${\cal E}_{\sigma,{\rm rel}}$ is not. Only in special cases can relative energy per particle ${\cal E}_{\sigma,{\rm rel}}$ be calculated from $f_\sigma$ at a particular time. Rather, only the increment $d{\cal E}_{\sigma,{\rm rel}}$ has an instantaneous physical meaning. This was pointed out in Ref.~\cite{Eu95}, and used as motivation to not employ relative entropy per particle because they sought a thermodynamic theory of irreversible processes. Our interpretation is distinctly different; we argue relative energy per particle not being a state variable reflects the physical consequence that changing the shape of $f_\sigma$ is typically history dependent.  Thus, a description retaining this history dependence is crucial for quantifying energy conversion into non-LTE internal moments.

Our results reveal that the standard treatment of energy conversion in Eq.~(\ref{eq:intenergyevolve}) needs to be expanded to accurately describe energy conservation when not in LTE. Since Eq.~(\ref{eq:intenergyevolve}) is equivalent to the first law of thermodynamics, we argue Eq.~(\ref{eq:firstlawfinal}) is its kinetic theory generalization, which we dub ``the first law of kinetic theory.'' 

A flow chart depicting energy conversion in non-LTE systems is in Fig.~\ref{fig:firstlaw}. Black arrows denote energy conversion contained in thermodynamics, namely conversion between heat, work, and internal energy, plus collisions. Red arrows are for relative energy and heat associated with non-LTE internal moments of $f_\sigma$. The dashed light blue arrow denotes coupling between relative energy and thermodynamic heat through the vector heat flux density and Pi-D.

We expect the results to be useful when $f_\sigma$ is reliably measured, such as PIC and Vlasov/Boltzmann plasma simulations and satellite observations \cite{Burch16,Pollock16}. Satellites measure $f_\sigma$ with spatio-temporal resolution sufficient to take gradients \cite{Shuster19,Shuster21} and compute kinetic entropy \cite{Argall22}. The theory may advance efforts using machine learning to parametrize kinetic corrections to transport terms in fluid models \cite{Laperre22}.  Generalizations of the present result may be useful beyond plasma physics, such as many body astrophysics \cite{Aarseth03}, micro- and nano-fluidics \cite{Karplus90,Schaller14}, and quantum entanglement \cite{Floerchinger20}.

There are limitations of the present work. Each restriction to the theory before Eq.~(\ref{eq:entropydensitydef}) could be relaxed. Relative energy describes energy conversion associated with all non-LTE internal moments, but does not identify which of the individual non-LTE internal moments contribute; it would be interesting to address this in future work, likely in context of recent theories of the velocity space cascade \cite{Servidio17} and/or Casimir invariants \cite{Zhdankin22}.  There are settings for which $f_{\sigma M}$ is not the appropriate reference for $f_\sigma$ \cite{LyndenBell67,Livadiotis_2018}; Ref.~\cite{Eu95} employs a more general reference $f_\sigma$ than we use here; it would be interesting to generalize the results for more general plasma-relevant forms.

\begin{acknowledgments}
The authors acknowledge helpful conversations with Amitava Bhattacharjee, Ned Flagg, the late Leo Golubovic, Greg Good, Colby Haggerty, David Levermore, Bill Matthaeus, Earl Scime, Mike Shay, Marc Swisdak, Eitan Tadmor, Thanos Tzavaras, and especially Art Weldon. The authors gratefully acknowledge support from NSF grant PHY-1804428 (PAC), NSF grant AGS-1602769 (PAC), NASA grant 80NSSC19M0146 (PAC), DOE grant DE-SC0020294 (PAC), NASA grant SV4-84017 (HL), NSF grant OIA-1655280 (HL), NASA grant 80GSFC19C0027 (HL), NASA grant 80NSSC20K1783 (HL), NASA grant 80NSSC21K0003 (HL), and NASA contract NNG04EB99C (MRA).  This research uses resources of the National Energy Research Scientific Computing Center (NERSC), a DOE Office of Science User Facility supported by the Office of Science of the US Department of Energy under Contract no.~DE-AC02-05CH11231.  Data used in Fig.~\ref{fig:entrpyFluxQuantities} is publicly available at https://doi.org/10.5281/zenodo.5847092, Ref.~\cite{zenodo22}.
\end{acknowledgments}


%

\pagebreak

\begin{center}
\large {\bf Supplemental Material for {\it Quantifying Energy Conversion in Higher Order Phase Space Density Moments in Plasmas}} \normalsize
\end{center}

\subsection{Derivation of Kinetic Entropy Evolution Equation}

The evolution equation for the kinetic entropy density $s_\sigma$ defined in Eq.~(2) is obtained by taking its partial time derivative and eliminating $\partial f_\sigma / \partial t$ using the Boltzmann equation \cite{Boltzmann1872},
\begin{equation}
\frac{\partial f_\sigma}{\partial t} + {\bf v} \cdot \nabla f_\sigma + \frac{{\bf F}_\sigma}{m_\sigma} \cdot \nabla_v f_\sigma = C[f],
    \label{eq:boltzeqn}
\end{equation}
where ${\bf F}_\sigma$ is the sum of any body forces, $\nabla_v$ is the velocity space gradient operator, and $C[f]$ is the inter- and intra-species collision operator, yielding \cite{Eu95,Jou10,Eyink18}
\begin{equation}
    \frac{\partial s_\sigma}{\partial t} + \nabla \cdot \bm{\mathcal{J}}_\sigma = \dot{s}_{\sigma,{\rm coll}},
    \label{eq:entdensev}
\end{equation}
where $\bm{\mathcal{J}}_\sigma$ is the kinetic entropy density flux
\begin{equation}
  \bm{\mathcal{J}}_\sigma = - k_{B} \int {\bf v} f_\sigma \ln\left(\frac{f_\sigma \Delta^3r_\sigma
    \Delta^3v_\sigma}{N_\sigma}\right) d^{3}v
    \label{eq:entdensflux}
\end{equation}
and $\dot{s}_{\sigma,{\rm coll}}$ is
the local time rate of change of kinetic entropy density through collisions,
\begin{equation}
  \dot{s}_{\sigma,{\rm coll}} = -k_B \int C[f] \ln \left( \frac{f_\sigma \Delta^3r_\sigma
    \Delta^3 v_\sigma}{N_\sigma} \right) d^3v. \label{sdotcoll}
\end{equation}
Note, there is no term containing body forces such as the electric and magnetic forces in Eq.~(\ref{eq:entdensev}) because the force term in Eq.~(\ref{eq:boltzeqn}) identically vanishes in deriving Eq.~(\ref{eq:entdensev}). An equivalent form of Eq.~(\ref{eq:entdensev}) comes from writing ${\bf v} = {\bf u}_\sigma + {\bf v}_\sigma^\prime$ in Eq.~(\ref{eq:entdensflux}), which implies $\bm{\mathcal{J}}_\sigma = s_\sigma {\bf u}_\sigma + \bm{\mathcal{J}}_{\sigma,{\rm th}}$, where the thermal kinetic entropy density flux $\bm{\mathcal{J}}_{\sigma,{\rm th}}$ is defined as
\begin{equation}
\bm{\mathcal{J}}_{\sigma,{\rm th}} = - k_{B} \int {\bf v}_\sigma^\prime f_\sigma \ln\left(\frac{f_\sigma \Delta^3r_\sigma \Delta^3v_\sigma}{N_\sigma}\right)d^3v. \label{eq:jthdef2}
\end{equation}
Then, Eq.~(\ref{eq:entdensev}) becomes \cite{Jou10}
\begin{equation}
    \frac{\partial s_\sigma}{\partial t} + \nabla \cdot (s_\sigma{\bf u}_\sigma + \bm{\mathcal{J}}_{\sigma,{\rm th}}) = \dot{s}_{\sigma,{\rm coll}}.
    \label{eq:entdensevcons}
\end{equation}
This equation is in the stationary (Eulerian) reference frame.  

Here, we manipulate Eq.~(\ref{eq:entdensevcons}) to derive an evolution equation for kinetic entropy per particle $s_\sigma/n_\sigma$ in a comoving (Lagrangian) frame [Eq.~(3)].  Using the convective derivative $d/dt = \partial/\partial t + {\bf u}_\sigma \cdot \nabla$ and dividing Eq.~(\ref{eq:entdensevcons}) by the density $n_\sigma$ gives
\begin{equation}
    \frac{1}{n_\sigma} \frac{d s_\sigma}{dt} + \frac{s_\sigma}{n_\sigma} (\nabla \cdot {\bf u}_\sigma) + \frac{\nabla \cdot \bm{\mathcal{J}}_{\sigma,{\rm th}}}{n_\sigma} = \frac{\dot{s}_{\sigma,{\rm coll}}}{n_\sigma}.
    \label{eq:entdensevcons2}
\end{equation}
Using the continuity equation $dn_\sigma/dt = - n_\sigma \nabla \cdot {\bf u}_\sigma$ (since $N_\sigma$ is assumed constant), we get
\begin{equation}
    \frac{1}{n_\sigma} \frac{d s_\sigma}{dt} - \frac{s_\sigma}{n_\sigma^2} \frac{dn_\sigma}{dt} + \frac{\nabla \cdot \bm{\mathcal{J}}_{\sigma,{\rm th}}}{n_\sigma} = \frac{\dot{s}_{\sigma,{\rm coll}}}{n_\sigma}.
    \label{eq:entdensevcons3}
\end{equation}
Finally, the two terms on the left are equal to $d(s_\sigma/n_\sigma)/dt$, which completes the derivation of Eq.~(3). 

We conclude this section with two important notes. First, Eq.~(2) is the ``Boltzmann'' form of kinetic entropy density $s_\sigma$.  For collisionless systems, any function of $f_\sigma$ is conserved, so other entropies could be defined \cite{Zhdankin22}. We choose the Boltzmann entropy because it reduces to the ideal fluid entropy density for a system in LTE and, for collisional systems, the total Boltzmann entropy $S_\sigma = \int s_\sigma d^3r$ obeys an H-theorem ($S_\sigma$ is non-decreasing in time) for a reasonably defined collision operator \cite{Boltzmann77}. Neither need be the case for other entropies. The present analysis may be redone for other entropies for future work.

Second, the approach we use remains valid even if there is an entropy source in the Boltzmann equation beyond collisions, such as due to boundaries of a finite domain. Such sources can lead to non-conservation of total kinetic entropy $S_\sigma = \int s_\sigma d^3r$ even in collisionless systems \cite{Grandy04}, but $s_\sigma$ is local in space and time and therefore remains well-defined.

\subsection{Derivation of Generalized Work Term}

Here, we derive Eq.~(5a). Dividing both sides of Eq.~(4a) by $n_\sigma$ and taking its total time derivative gives
\begin{equation}
    \frac{d}{dt}\left(\frac{s_{\sigma p}}{n_\sigma}\right) = - \frac{k_B}{n_\sigma} \frac{dn_\sigma}{dt}.
\end{equation}
A brief derivation reveals this is equivalent to 
\begin{equation}
    \frac{d}{dt}\left(\frac{s_{\sigma p}}{n_\sigma}\right) = k_B n_\sigma \frac{d(1/n_\sigma)}{dt}. \label{eq:ddtsponderive}
\end{equation}
Defining $V_\sigma = 1/n_\sigma$ as the volume per particle, and using ${\cal P}_\sigma = n_\sigma k_B {\cal T}_\sigma$, the previous equation is equivalent to
\begin{equation}
    \frac{d}{dt}\left(\frac{s_{\sigma p}}{n_\sigma}\right) = \frac{1}{{\cal T}_\sigma} \frac{d{\cal W}_\sigma}{dt}, 
\end{equation}
where $d{\cal W}_\sigma = {\cal P}_\sigma dV_\sigma = {\cal P}_\sigma d(1/n_\sigma)$ is the non-LTE generalization of the work per particle done by the system.  

To physically interpret this, note $s_{\sigma p}$ is associated with the number of permutations of particles in position space that produce the same macrostate without concern for their velocity \cite{Liang19}. The argument of the natural logarithm in $s_{\sigma p}/n_\sigma = -k_B \ln (n_\sigma \Delta^3r_\sigma/N_\sigma)$ is always between 0 and 1, so $s_{\sigma p}/n_\sigma$ is non-negative and is a strictly decreasing function of $n_\sigma$.  Thus, local compression ($d{\cal W}_\sigma = {\cal P}_\sigma dV_\sigma <0$) increases $n_\sigma$ and decreases $s_{\sigma p}/n_\sigma$ [{\it i.e.,} $d(s_{\sigma p}/n_\sigma)/dt < 0$], while local expansion ($d{\cal W}_\sigma = {\cal P}_\sigma dV_\sigma > 0$) decreases $n_\sigma$ and increases $s_{\sigma p}/n_\sigma$ [{\it i.e.,} $d(s_{\sigma p}/n)/dt > 0$].

\subsection{Derivation of Generalized Energy Term}

We next derive Eq.~(5b). We decompose the velocity space kinetic entropy density $s_{\sigma v}$ in Eq.~(4b) as
$s_{\sigma v} = s_{\sigma v,{\cal E}} + s_{\sigma v,{\rm rel}}$, where
\begin{eqnarray}
s_{\sigma v,{\cal E}} & = & -k_B \int f_\sigma \ln \left(\frac{f_{\sigma M} \Delta^3v_\sigma}{n_\sigma}\right) d^3v, \label{eq:svenergy} \\
s_{\sigma v,{\rm rel}} & = & -k_B \int f_\sigma \ln \left(\frac{f_\sigma}{f_{\sigma M}}\right) d^3v, \label{eq:svrel2}
\end{eqnarray}
where $f_{\sigma M}$ is the Maxwellianized distribution of $f_\sigma$ defined in Eq.~(8). The relative entropy is related to the Kullback-Leibler divergence \cite{Kullback51} from information theory which is a measure of the statistical difference between a two probability distributions, and has been extensively used in a variety of fields, such as statistical mechanics, applied mathematics, chemistry, biology, quantum information theory, and economics \cite{Jaynes63,Grad65,Diperna79,Dafermos79,Eu95,Vedral02,Robertson05,Tzavaras05,Shell08,Berthelin09,Baez16}.  Substituting Eq.~(8) into Eq.~(\ref{eq:svenergy}) and carrying out straight-forward manipulations gives
\begin{equation}
\frac{s_{\sigma v,{\cal E}}}{n_\sigma} = \frac{3}{2} k_{B} \left[ 1 + \ln \left(\frac{2 \pi k_B {\cal T}_\sigma}{m_\sigma (\Delta^3v)^{2/3}}\right) \right]. \label{eq:scvovern3} 
\end{equation}
Its Lagrangian time derivative immediately gives
\begin{equation}
\frac{d}{dt} \left( \frac{s_{\sigma v,{\cal E}}}{n_\sigma} \right) = \frac{1}{{\cal T}_\sigma} \frac{d{\cal E}_{\sigma,{\rm int}}}{dt}, \label{eq:ddtsceq2}
\end{equation}
where $d{\cal E}_{\sigma,{\rm int}} = (3/2) k_B d{\cal T}_\sigma$ is the increment in internal energy per particle. This reproduces Eq.~(5b). Thus, $d(s_{\sigma v,{\cal E}}/n_\sigma)/dt > 0$ implies the effective temperature increases, while $d(s_{\sigma v,{\cal E}}/n_\sigma)/dt < 0$ implies the effective temperature decreases.  Physically, $s_{\sigma v}$ is associated with the number of permutations of particles of different velocities in a given position in phase space that produces the same macrostate \cite{Liang19}.

\subsection{Derivation of Generalized Heat Term}

We now derive Eq.~(5c). We find it is advantageous to decompose $\nabla \cdot \bm{\mathcal{J}}_{\sigma,{\rm th}}$ using Eq.~(\ref{eq:jthdef2}) as
\begin{equation}
      \nabla \cdot \bm{\mathcal{J}}_{\sigma,{\rm th}} = (\nabla \cdot \bm{\mathcal{J}}_{\sigma,{\rm th}})_{{\cal W}} + (\nabla \cdot \bm{\mathcal{J}}_{\sigma,{\rm th}})_{{\cal E}} + (\nabla \cdot \bm{\mathcal{J}}_{\sigma,{\rm th}})_{{\rm rel}}, \label{eq:jsigthdecomp}
\end{equation}
where
\begin{subequations}
\begin{eqnarray}
  (\nabla \cdot \bm{\mathcal{J}}_{\sigma,{\rm th}})_{{\cal W}} & = & - k_{B} \int (f_\sigma {\bf v}_\sigma^\prime) \cdot \nonumber \\ & & \nabla \left[\ln\left(\frac{f_\sigma \Delta^3r_\sigma \Delta^3v_\sigma}{N_\sigma}\right) \right] d^{3} v, \label{eq:divjthwdef} \\
  (\nabla \cdot \bm{\mathcal{J}}_{\sigma,{\rm th}})_{{\cal E}} & = & - k_{B} \int \left[\frac{}{} \nabla \cdot \left({\bf v_\sigma}^\prime f_\sigma\right)\right] \nonumber \\ & & \ln\left(\frac{f_{\sigma M} \Delta^3r_\sigma \Delta^3v_\sigma}{N_\sigma}\right) d^{3} v, \label{eq:divjthedef}
\end{eqnarray}
\end{subequations}
and $(\nabla \cdot \bm{\mathcal{J}}_{\sigma,{\rm th}})_{{\rm rel}}$ is defined in Eq.~(7).  The latter has equivalent forms of
\begin{eqnarray*}
  (\nabla \cdot \bm{\mathcal{J}}_{\sigma,{\rm th}})_{{\rm rel}} & = & - \nabla \cdot ({\bf u}_\sigma s_{\sigma v,{\rm rel}}) \nonumber \\ & & - k_B \int \nabla \cdot({\bf v}f_\sigma) \ln \left( \frac{f_\sigma}{f_{\sigma M}} \right) d^3v, \\
    & = & - s_{\sigma v,{\rm rel}} (\nabla \cdot {\bf u}_\sigma) \nonumber \\ & & - k_B \int ({\bf v}_\sigma^\prime \cdot \nabla f_\sigma) \ln \left( \frac{f_\sigma}{f_{\sigma M}} \right) d^3v, 
\end{eqnarray*}
which may be useful in applications depending on which quantities are easiest to measure in a given system.

We first treat $(\nabla \cdot \bm{\mathcal{J}}_{\sigma,{\rm th}})_{{\cal W}}/n_\sigma$.  The gradient of the term in brackets in Eq.~(\ref{eq:divjthwdef}) is $(1/f_\sigma) \nabla f_\sigma$.  Using ${\bf v}_\sigma^\prime = {\bf v} - {\bf u}_\sigma$, straight-forward manipulations give
\begin{equation}
\frac{(\nabla \cdot \bm{\mathcal{J}}_{\sigma,{\rm th}})_{{\cal W}}}{n_\sigma} = - k_{B} \nabla \cdot {\bf u}_\sigma = - k_B n_\sigma \frac{d(1/n_\sigma)}{dt}, \label{eq:divjthw}
\end{equation}
where we use the continuity equation to eliminate $\nabla \cdot {\bf u}_\sigma$. Therefore, this term describes the non-LTE generalization of heating associated with compression or expansion, with the same form as $d(s_{\sigma p}/n_\sigma)/dt$ in Eq.~(\ref{eq:ddtsponderive}) but with the opposite sign. This motivates our use of the ${\cal W}$ subscript.

Turning to $(\nabla \cdot \bm{\mathcal{J}}_{\sigma,{\rm th}})_{{\cal E}}/n_\sigma$, we use Eq.~(8) to write Eq.~(\ref{eq:divjthedef}) as
\begin{eqnarray*}
& & \frac{(\nabla \cdot \bm{\mathcal{J}}_{\sigma,{\rm th}})_{{\cal E}}}{n_\sigma} = - \frac{k_{B}}{n_\sigma} \int \nabla \cdot \left({\bf v}_\sigma^\prime f_\sigma \right) \\ & & \left\{ \ln \left[\left(\frac{n_\sigma \Delta^3r_\sigma \Delta^3v_\sigma}{N_\sigma}\right)\left( \frac{m_\sigma}{2\pi k_B {\cal T}_\sigma} \right)^{3/2} \right] - \frac{m_\sigma v_\sigma^{\prime 2}}{2k_B {\cal T}_\sigma} \right\} d^{3} v.
\end{eqnarray*}
The term in square brackets is independent of ${\bf v}$ and hence comes out of the integral, and the remaining part of that integral is $\int \nabla \cdot ({\bf v}_\sigma^\prime f_\sigma) d^3v = 0$. Manipulations of the remaining term after integration by parts gives
\begin{eqnarray}
\frac{(\nabla \cdot \bm{\mathcal{J}}_{\sigma,{\rm th}})_{{\cal E}}}{n_\sigma} & & = \frac{\nabla \cdot \left( {\bf q}_\sigma / {\cal T}_\sigma \right)}{n_\sigma} - \nonumber \\
& & \frac{m_\sigma}{2 n_\sigma} \int f_\sigma {\bf v}_\sigma^\prime \cdot \left[\frac{\nabla v_\sigma^{\prime 2}}{{\cal T}_\sigma} - \frac{v_\sigma^{\prime 2} \nabla {\cal T}_\sigma}{{\cal T}_\sigma^2} \right] d^{3} v,
\end{eqnarray}
where ${\bf q}_\sigma$ is the vector heat flux density defined after Eq.~(1).
Using index notation and the Einstein summation convention,
\[
{\bf v}_\sigma^\prime \cdot \nabla v_\sigma^{\prime 2} = v_{\sigma j}^\prime \frac{\partial (v_{\sigma k}^\prime v_{\sigma k}^\prime)}{\partial r_j} = 2 v_{\sigma j}^\prime v_{\sigma k}^\prime \frac{\partial v_{\sigma k}^\prime}{\partial r_j} = -2 v_{\sigma j}^\prime v_{\sigma k}^\prime \frac{\partial u_{\sigma k}}{\partial r_j},
\]
where we use $\partial v_{\sigma k}^\prime/\partial r_j = \partial (v_k-u_{\sigma k})/\partial r_j = -\partial u_{\sigma k}/\partial r_j$ in the last equality.  Carrying out the remaining integrals and simplifying gives
\begin{equation}
\frac{(\nabla \cdot \bm{\mathcal{J}}_{\sigma,{\rm th}})_{{\cal E}}}{n_\sigma} = \frac{\nabla \cdot {\bf q}_\sigma}{n_\sigma {\cal T}_\sigma} + \frac{({\bf P}_\sigma \cdot \nabla) \cdot {\bf u}_\sigma}{n_\sigma {\cal T}_\sigma}. \label{eq:divjtheon}
\end{equation}
Comparing the right hand side of Eq.~(\ref{eq:divjtheon}) with Eq.~(1), we see both terms appear directly in the internal energy equation with the opposite sign, motivating the choice of the subscript ${\cal E}$.  This term describes heat per particle that changes only the effective temperature. A negative value of $(\nabla \cdot \bm{\mathcal{J}}_{\sigma,{\rm th}})_{\cal E}/n_\sigma$ drives increasing ${\cal T}_\sigma$, and a positive value drives decreasing ${\cal T}_\sigma$.

A consequence of Eq.~(\ref{eq:divjthw}) is that 
$(\nabla \cdot \bm{\mathcal{J}}_{\sigma,{\rm th}})_{{\cal W}} / n_\sigma = - ({\cal P}_\sigma / n_\sigma {\cal T}_\sigma) \nabla \cdot {\bf u}_\sigma$, so
\begin{eqnarray}
\frac{(\nabla \cdot \bm{\mathcal{J}}_{\sigma,{\rm th}})_{{\cal W}}}{n_\sigma} + \frac{(\nabla \cdot \bm{\mathcal{J}}_{\sigma,{\rm th}})_{{\cal E}}}{n_\sigma} & = & \frac{\nabla \cdot {\bf q}_\sigma}{n_\sigma {\cal T}_\sigma} + \frac{\Pi_{\sigma,jk} {\cal D}_{\sigma,jk}}{n_\sigma {\cal T}_\sigma} \nonumber \\ & = & - \frac{1}{{\cal T}_\sigma} \frac{d{\cal Q}_\sigma}{dt}, \label{eq:dqsigmadt}
\end{eqnarray}
where we use the known decomposition $({\bf P}_\sigma \cdot \nabla) \cdot {\bf u}_\sigma = {\cal P}_\sigma (\nabla \cdot {\bf u}_\sigma) + \Pi_{\sigma,jk} {\cal D}_{\sigma,jk}$, with $\Pi_{\sigma,jk} = P_{\sigma,jk} - {\cal P}_\sigma \delta_{jk}$ being elements of the deviatoric pressure tensor $\bm{\Pi}$, ${\cal D}_{\sigma,jk} = (1/2) (\partial u_{\sigma j}/\partial r_k + \partial u_{\sigma k} / \partial r_j) - (1/3) \delta_{jk} (\nabla \cdot {\bf u}_\sigma)$ being elements of the traceless symmetric strain rate tensor $\bm{{\cal D}}$, and $\delta_{jk}$ being the Kroenecker delta \cite{Jou10,Yang17}. This derivation provides the expression given in Eq.~(5c).

\subsection{Derivation of Relative Energy Per Particle Rate for a Bi-Maxwellian Plasma}

Here we derive the rate of relative energy per particle change $d{\cal E}_{\sigma,{\rm rel}}/dt$ for a bi-Maxwellian initial phase space density. Consider a purely collisionless system in which the initial $f_\sigma$ is bi-Maxwellian with a converging bulk flow ${\bf u}({\bf r},t)$.  We define $z$ as the parallel direction $\parallel$, $x$ and $y$ as perpendicular $\perp$ directions, and $T_\perp$ and $T_\parallel$ as the uniform temperatures in the $\perp$ and $\parallel$ directions. The initial phase space density is \begin{eqnarray}
f_{biM} & = & n \left(\frac{m}{2\pi k_B T_\perp}\right) \left(\frac{m}{2\pi k_B T_\parallel}\right)^{1/2} \nonumber \\ & & e^{-m[(v_x - u_x)^2 + (v_y - u_y)^2]/2 k_B T_\perp} e^{-m(v_{z} - u_z)^2/2 k_B T_\parallel}, 
\end{eqnarray}
where $n$ is the initial number density and the constituent mass is $m$. The Maxwellianized distribution for this system has the form of Eq.~(8) with effective temperature ${\cal T} = (2T_\perp + T_\parallel)/3$. Then, 
\begin{eqnarray*}
\ln\left( \frac{f_{biM}}{f_{M}} \right) & = & \ln\left( \frac{{\cal T}^{3/2}}{T_\perp T_\parallel^{1/2}} \right) - \frac{m(v_x^{\prime 2} + v_y^{\prime 2})}{2k_B} \left[\frac{T_\parallel-T_\perp}{T_\perp(2T_\perp + T_\parallel)} \right] \\ & & - \frac{m v_z^{\prime 2}}{2k_B}  \left[\frac{2(T_\perp-T_\parallel)}{T_\parallel(2T_\perp + T_\parallel)} \right],
\end{eqnarray*}
and a straight-forward derivation using Eq.~(6) yields
\begin{eqnarray}
\frac{s_{v,{\rm rel}}}{n} & = & -\frac{3}{2} k_B \ln \left[ \frac{2}{3}\left(\frac{T_\perp}{T_\parallel}\right)^{1/3} + \frac{1}{3}\left(\frac{T_\parallel}{T_\perp}\right)^{2/3} \right].
\end{eqnarray}
The Lagrangian time derivative of this equation, after some algebra, gives
\begin{equation}
\frac{d}{dt} \left(\frac{s_{v,{\rm rel}}}{n} \right) = k_B \left( \frac{T_\parallel - T_\perp}{2T_\perp+T_\parallel} \right) \left( \frac{1}{T_\perp} \frac{dT_\perp}{dt} - \frac{1}{T_\parallel} \frac{dT_\parallel}{dt}\right). \label{eq:ddtsrelbim}
\end{equation}
The thermal evolution equations in a collisionless gyrotropic system, which follow directly from the second parallel and perpendicular moments of the collisionless Boltzmann equation, are written in terms of parallel and perpendicular pressures as \cite{Chew56,Hesse92}
\begin{subequations}
\begin{eqnarray}
\frac{dP_\parallel}{dt} + P_\parallel \nabla \cdot {\bf u} + 2 P_\parallel \left[ {\bf \hat{z}} \left( {\bf \hat{z}} \cdot \nabla\right)\right] \cdot {\bf u} & = & 0, \\
\frac{dP_\perp}{dt} + 2 P_\perp \nabla \cdot {\bf u} - P_\perp \left[ {\bf \hat{z}} \left( {\bf \hat{z}} \cdot \nabla\right)\right] \cdot {\bf u} & = & 0, 
\end{eqnarray}
\end{subequations}
where $P_\perp = nk_B T_\perp$ and $P_\parallel = nk_B T_\parallel$. Substituting these into Eq.~(\ref{eq:ddtsrelbim}) gives  
\begin{equation}
\frac{d}{dt} \left(\frac{s_{v,{\rm rel}}}{n} \right) = k_B \left( \frac{T_\parallel - T_\perp}{2T_\perp+T_\parallel} \right) \left(-\nabla_\perp \cdot {\bf u}_\perp + 2 \frac{\partial u_z}{\partial z}\right), \label{eq:ddtsrelbim2}
\end{equation}
where ${\bf u}_\perp = {\bf u} - {\bf \hat{z}} u_z$. Finally, using Eq.~(9a) to relate this to $d{\cal E}_{\sigma,{\rm rel}}/dt$ gives 
\begin{equation}
\frac{d{\cal E}_{\sigma,{\rm rel}}}{dt} = \frac{1}{3} k_B (T_\parallel - T_\perp) \left(-\nabla_\perp \cdot {\bf u}_\perp + 2 \frac{\partial u_z}{\partial z}\right). \label{eq:ddtsrelbim3}
\end{equation}

To interpret this result physically, suppose $T_\parallel > T_\perp$.  First consider a bulk flow profile ${\bf u} = {\bf u}_\perp$ that is isotropically converging in the $xy$ plane. Compression leads to heating, but only in the perpendicular direction \cite{Cassak_PiD1_2022}. Thus, $f_\sigma$ becomes more Maxwellian.   From Eq.~(\ref{eq:ddtsrelbim3}), both $T_\parallel - T_\perp$ and the bulk velocity term are positive, so $d{\cal E}_{\sigma,{\rm rel}}/dt > 0$, consistent with energy going into higher order moments making $f_\sigma$ more Maxwellian. Now consider converging bulk flow in the $z$ direction. The compression heats the distribution in the parallel direction, so $f_\sigma$ becomes more elongated in the parallel direction, {\it i.e.,} less Maxwellian. From Eq.~(\ref{eq:ddtsrelbim3}), $T_\parallel - T_\perp$ is positive but the bulk velocity term is negative, so this evolution away from Maxwellianity is consistent with $d{\cal E}_{\sigma,{\rm rel}}/dt$ being negative. This example illustrates general features: $d{\cal E}_{\sigma,{\rm rel}}/dt > 0$ is associated with energy conversion making higher moments of $f_\sigma$ evolve to become more Maxwellian, and vice-versa.

\subsection{Numerical Simulation Methodology}
\label{sec:simsetup}

Details of the simulation in addition to what follows are available in Ref.~\cite{Barbhuiya_PiD3_2022}.  We use the massively parallel particle-in-cell code {\tt p3d} \cite{zeiler:2002} to perform simulations that are 3D in velocity-space and 2.5D in position-space.  Periodic boundary conditions are used in both spatial directions. The code uses the relativistic Boris particle stepper \cite{birdsall91a} for stepping particles forward in time, while the trapezoidal leapfrog method \cite{guzdar93a} is utilized for stepping electromagnetic fields forward in time.  The fields have a time-step half of that of the particles. The multigrid method \cite{Trottenberg00} is used to clean the electric field ${\bf E}$ to enforce Poisson's equation every 10 particle time-steps.

Simulation results are presented in normalized units. The reference magnetic field $B_0$ is the initial asymptotic magnetic field strength.  The reference number density $n_0$ is the peak current sheet number density minus the asymptotic background number density. Length scales are normalized to the ion inertial scale $d_{i0} = c/\omega_{pi0}$, where $\omega_{pi0} = (4 \pi n_0 q_i^2 /m_i)^{1/2}$ is the ion plasma frequency (in cgs units), $q_i$ is the ion charge, $m_i$ is the ion mass, and $c$ is the speed of light. Velocities are normalized to the Alfv\'en speed $c_{A0} = B_0/(4 \pi m_i n_0)^{1/2}$.  Times are normalized to the inverse ion cyclotron frequency $\Omega_{ci0}^{-1}= (q_i B_{0} / m_{i} c)^{-1}$. Temperatures are normalized to $m_i c_{A0}^2/k_B$.  Current densities are normalized to $cB_0 / 4 \pi d_{i0}$. Reduced phase space densities, with one velocity dimension integrated out, are normalized to $n_0/c_{A0}^2$. Energy per particle conversion rates are given in units of $\Omega_{ci0} B_0^2 / 4 \pi n_0$. 

The initial condition has two oppositely directed current sheets with drifting Maxwellian initial distributions. The magnetic field profile is a double tanh with no initial out-of-plane (guide) magnetic field. The current sheet thickness is $w_0 = 0.5$, the background density is $n_{up}=0.2$, and the electron and ion temperatures are 1/12 and 5/12, respectively. A magnetic perturbation of amplitude 0.05 seeds an X-line/O-line pair in each of the two current sheets.  The simulation system size is $L_x \times L_y = 12.8 \times 6.4$, where $x$ and $y$ are the outflow and inflow directions, respectively. The speed of light $c$ is 15 and the electron to ion mass ratio is $m_e/m_i = 0.04$. There are $1024 \times 512$ grid cells initialized with 25,600 weighted particles per grid (PPG), which is chosen to be very large to decrease particle noise. The grid-length $\Delta$ in both directions is 0.0125, which is smaller than the smallest length scale which is the electron Debye length of 0.0176. The time-step $\Delta t$ is 0.001, which is smaller than the smallest time scale which is the inverse of electron plasma frequency of 0.012.  Our choice of these numerical parameters results in a total energy increase by only 0.022\% by $t=14$.

All plots display data from only the lower current sheet at time $t = 13$, when the rate of reconnection is increasing most rapidly in time. To reduce PIC noise for all quantities plotted other than phase space densities, the raw quantities are recursively smoothed four times over a width of four cells, then any temporal or spatial derivatives are carried out, and then the results are again smoothed recursively four times over four cells. For temporal derivatives, the presented data is calculated from a finite difference between $t = 12$ and $14$ (on ion cyclotron time scales).  The results are compared to those obtained from a finite difference between $t = 12.96$ and $13.04$ (electron time scales), and the results are found to differ by less than 5\%; this change is deemed inconsequential for the present purposes.

Kinetic entropy is calculated in the simulations employing the implementation from Ref.~\cite{Argall22}. Optimization of the velocity-space grid \cite{Liang20b} is done by checking the agreement between the kinetic entropy density for electrons $s_e$ calculated by the simulation for various $\Delta v_e$ and the theoretical value at $t=0$. We find an optimal $\Delta v_e$ of 1.33 which leads to an initial agreement to within $\pm 1\%$ in the upstream region and $\pm 3\%$ at the center of the current sheet. For plots of reduced electron phase space densities, we use a domain of size $0.0625 \times 0.0625$ centered at the location of interest. Particles are binned with a velocity space bin of size 0.1 in all velocity directions.

\subsection{Additional Comments on the Relation to Previous Work}

Here, we put our result in context of previous work on related topics. 
\begin{itemize}
\item {\bf Energy Conversion in $\delta f_\sigma$ Kinetic Theory and Gyrokinetics:} We first compare the present work with previous work on energy conversion in linearized kinetic theory and gyrokinetics \cite{Hallatschek04,Howes06,Schekochihin09,Tatsuno09}. Consider a linear expansion of the phase space density about its Maxwellianized distribution, so that $f_\sigma = f_{\sigma M} + \delta f_\sigma$, and $\delta f_\sigma \ll f_{\sigma M}$.  A straight-forward calculation using Eq.~(\ref{eq:svrel2}) reveals that the linearized relative entropy $\delta s_{\sigma v,{\rm rel}}$ is
\begin{equation}
\delta s_{\sigma v,{\rm rel}} \simeq -k_B \int \frac{(\delta f_\sigma)^2}{2 f_{\sigma M}} d^3v. \label{eq:svrellinear}
\end{equation}
In linear theory, the density and temperature in $f_{\sigma M}$ are their equilibrium values, which we call $n_{\sigma 0}$ and $T_{\sigma 0}$, respectively. Then, the linearized equation describing the relative energy increment using Eq.~(9a) is
\begin{equation}
\frac{d{\cal E}_{\sigma,{\rm rel}}}{dt} \simeq T_{\sigma 0}\frac{d(s_{\sigma v,{\rm rel}}/n_{\sigma 0})}{dt}. \label{eq:derellinear}
\end{equation}
Since the equilibrium temperature does not change to low order in linear theory, $n_{\sigma 0}$ and $T_{\sigma 0}$ are constant in time, so integrating Eq.~(\ref{eq:derellinear}) in time gives
\begin{eqnarray}
\delta{\cal E}_{\sigma,{\rm rel}} & \simeq & \frac{T_{\sigma 0}}{n_{\sigma 0}} \delta s_{\sigma v,{\rm rel}} \nonumber \\ & \simeq & - \frac{k_B T_{\sigma 0}}{n_{\sigma 0}} \int \frac{(\delta f_\sigma)^2}{2 f_{\sigma M}} d^3v, \label{eq:linrelent}
\end{eqnarray}
where we use Eq.~(\ref{eq:svrellinear}).

In comparison, the free energy in a $\delta f_\sigma$ linearized thermodynamic approach   \cite{Hallatschek04} [$\kappa_{\rm int}$ in their Eq.~(7)] was derived to be
\begin{equation}
\kappa_{\rm int} = k_B T_{0} \int \frac{(\delta f)^2}{2 f_{\sigma M}} d^3v \label{eq:kappaint}
\end{equation}
and in a gyrokinetic analysis of energy conversion \cite{Howes06,Schekochihin09}, the comparable term from the free energy \{the first term of $W$ in Eq.~(74) from Ref.~\cite{Schekochihin09}\} is
\begin{equation}
    W = \int d^3r \sum_\sigma \int k_B T_{0 \sigma} \frac{(\delta f)^2}{2 f_{\sigma M}} d^3v.
\end{equation}

Clearly, the linearized relative energy per particle $\delta {\cal E}_{\sigma,{\rm rel}}$ in Eq.~(\ref{eq:linrelent}) is related to the free energy in the $\delta f_\sigma$ thermodynamic and the gyrokinetic approaches. In particular, $\kappa_{\rm int} = - n_{\sigma 0} \delta {\cal E}_{\sigma,{\rm rel}}$ and $W = - \int d^3r \sum_\sigma n_{\sigma 0} \delta {\cal E}_{\sigma,{\rm rel}}$. [Note, the relative entropy term differs from the nonlinear term used in Ref.~\cite{Cerri18} that reproduces Eq.~(\ref{eq:kappaint}) when linearized; theirs is related to $\bar{M}_{KP}$ rather than the relative energy term.] The sign difference is a result of $\delta {\cal E}_{\sigma,{\rm rel}}$ measuring the energy going into the random energy of the particles, while $\kappa_{\rm int}$ and $W$ describe energy going into the bulk flow energy and magnetic fields from the particles. Thus, the present work is consistent with previous work, and generalizes these linear approaches for phase space densities arbitrarily far from LTE.

\item {\bf Previous Schematics of Energy Conversion:} We now put the sketch of energy conversion in Fig.~3 in the context of previous sketches about energy conversion in plasmas.  It is similar to Fig.~1 in Ref.~\cite{Yang17}, except theirs is averaged over a closed or periodic domain so the heat flux does not contribute, theirs includes conversion into bulk kinetic energy and electromagnetic energy which are omitted from the present treatment for simplicity, and ours includes collisions. The key difference is the additional energy conversion channel associated with relative energy and heat that arise from our analysis as another possible energy conversion channel.

Another related sketch is Fig.~4 in Ref.~\cite{Howes18}, which describes energy conversion in weakly collisional turbulent plasmas. There, electromagnetic fields play a key role in converting energy to non-thermal (non-LTE) energy in the plasma, which ultimately produce irreversible dissipation through the collisions. The present work treats only internal moments of the phase space density, which formally has only indirect input from body forces [which, for example, do not appear in Eq.~(1)]. Thus, our result is in many ways complementary to the research done on the field-particle correlation \cite{Klein16}.  It would be interesting and important to unite the two approaches in future work.

\item {\bf The Velocity Space Cascade and Hermite Expansions of $f_\sigma$:} An important approach that has previously been used to study non-LTE energy conversion is to take a local phase space density and expand the velocity space part in Hermite polynomials \cite{Servidio17,Pezzi18,Pezzi19}. The coefficients in the expansion provide information about how non-Maxwellian the system is at that location in space and time. In a weakly collisional or collisionless system, many phase space densities develop sharp structures in velocity space, which shows up as a cascade of power into the higher order coefficients in the expansion.

It would be tempting to associate the power in non-LTE modes, called the enstrophy in Ref.~\cite{Servidio17}, with the relative energy per particle in the present analysis, but this association is not possible. The reason is that the enstrophy is a local quantity that can be calculated for any phase space density, but the relative energy per particle is history dependent, so only changes to it can be uniquely determined from the local phase space density at a particular time. A phase space density becoming more non-Maxwellian has an increase in enstrophy, while it corresponds to a decrease in the relative energy per particle because the Maxwellian is the maximum entropy state.  While associating the two approaches in this manner is therefore not possible, we do believe there are links between the two approaches which will be pursued in future studies.

\item {\bf Energy Conversion Using Other Entropies:} Recent work quantified non-LTE effects using non-Boltzmann entropies for collisionless plasmas \cite{Zhdankin22,Zhdankin22b}. In Ref.~\cite{Zhdankin22}, energy conversion was parametrized by moments of integer powers of $f_\sigma$, which are invariants in collisionless systems.  In Ref.~\cite{Zhdankin22b}, it was shown that power law entropies are well-suited for describing power law tails during non-thermal particle acceleration. As pointed out there, these terms provide information about the shape of the phase space density, so there are some similarities about the aims of the two studies despite their different approaches.

The formulation here using the Boltzmann entropy is related to these invariants, as an expansion of the natural logarithm in powers of $f_\sigma$ inside the kinetic entropy density $s_\sigma = -k_B \int f_\sigma \ln (f_\sigma \Delta^3r_\sigma \Delta^3v_\sigma / N_\sigma) d^3v$ yields integrals over all integer powers of $f_\sigma$, as done in Ref.~\cite{Zhdankin22}. Consequently, the form derived here based on Boltzmann entropy without expanding the natural logarithm automatically contains the information about all of the power law invariants for collisionless systems. Ref.~\cite{Zhdankin22} is important for identifying how the energy is contained in different individual invariants, which is not possible in the present formulation. However, our results can readily be used for collisional systems even though powers of $f_\sigma$ are no longer invariants.

\item {\bf Extended Irreversible Thermodynamics (EIT)}:  EIT begins with the kinetic entropy evolution equation [Eq.~(\ref{eq:entdensev})] and employs a perturbative expansion of $f_\sigma$, and the terms of higher order represent corrections to the first law of thermodynamics.  This is very important because the corrections are in terms of fluid moments of $f_\sigma$, so a direct measurement of $f_\sigma$ is not necessary. The advantage of the present analysis is that all internal moments are retained, so there is no need to be near LTE. 

We also point out that the phase space density $f_\sigma$ inside the natural logarithm in the general expression for $\bm{\mathcal{J}}_{\sigma,{\rm th}}$ [Eq.~(\ref{eq:jthdef2})] is expanded about the Maxwellianized distribution $f_{\sigma M}$ in EIT.  The lowest order term in this expansion is \cite{Jou10}
\begin{equation}
\bm{\mathcal{J}}_{\sigma,q} = -k_B \int {\bf v}_\sigma^\prime f_\sigma \ln\left(\frac{f_{\sigma M} \Delta^3r_\sigma \Delta^3 v_\sigma}{N_\sigma}\right) d^3v. \label{eq:jsqdef}
\end{equation}
A brief derivation using Eq.~(8) reveals that $\bm{\mathcal{J}}_{\sigma,q} = {\bf q}_\sigma/{\cal T}_\sigma$.  In the present study, instead of decomposing $f_\sigma$ inside $\bm{\mathcal{J}}_{\sigma,{\rm th}}$, we decompose $f_\sigma$ inside $\nabla \cdot \bm{\mathcal{J}}_{\sigma,{\rm th}}$ as Eq.~(\ref{eq:jsigthdecomp}).  The difference here is that $\bm{\mathcal{J}}_{\sigma,q} = {\bf q}_\sigma/{\cal T}_\sigma$ from Eq.~(\ref{eq:jsqdef}), so $\nabla \cdot \bm{\mathcal{J}}_{\sigma,q}$ contains both a $(\nabla \cdot {\bf q}_\sigma) / {\cal T}_\sigma$ term and a $-({\bf q}_\sigma \cdot \nabla {\cal T}_\sigma) / {\cal T}_\sigma^2$ term.  The latter term is included as an entropy source term in the fluid form of EIT \cite{Jou10}. Eq.~(\ref{eq:divjtheon}) reveals that $-({\bf q}_\sigma \cdot \nabla {\cal T}_\sigma)/{\cal T}_\sigma^2$ vanishes exactly when all orders of non-LTE terms are retained so that it should not be retained. 

\item {\bf Quantum Statistical Mechanics:} There are similarities and differences of our results with a recent independent analysis showing that the quantum first law of thermodynamics can be obtained from the quantum relative entropy \cite{Floerchinger20}.  In the classical limit, the density matrix $\rho$ is analogous to the distribution function $f_\sigma / n_\sigma$ \cite{Sakurai94}.  The maximally mixed state $\sigma_m$, which has the highest entropy, is analogous to the Maxwellianized distribution function $f_{\sigma M}/n_\sigma$.  The von Neumann entropy $S(\rho) = -{\rm tr}[\rho \ln \rho]$ \cite{vonNeumann27} is decomposed as $S(\rho) = S_{{\rm cross}}(\rho) - S_{{\rm rel}}(\rho)$, where $S_{{\rm cross}}(\rho) = - {\rm tr}[\rho \ln \sigma_{m}]$ is the cross-entropy and $S_{{\rm rel}} = {\rm tr}[\rho \ln \rho - \rho \ln \sigma_m] = -S(\rho) + S_{{\rm cross}}(\rho)$ is the relative entropy \cite{Floerchinger20}.  This is similar to the decomposition done here for the velocity space kinetic entropy per particle, so $S_{{\rm cross}}(\rho)$ is analogous to $s_{\sigma v,{\cal E}}/n_\sigma$ [Eq.~(\ref{eq:svenergy})] and $S_{{\rm rel}}(\rho)$ is analogous to $-s_{\sigma v,{\rm rel}}/n_\sigma$ [Eq.~(6)]. In Ref.~\cite{Floerchinger20}, the volume of the system was kept fixed for simplicity, so there was no term analogous to the position space entropy term in our analysis. Including this term, which gives rise to work in the classical case, is very straight-forward; indeed, it appears automatically when the phase space density $f_\sigma$ is employed instead of the distribution function $f_\sigma/n_\sigma$. Undoubtedly the quantum statistical mechanical approach can be generalized to include work done on the system using open quantum mechanics \cite{Lidar20}.  

For the classical case presented here, the physical interpretation of the terms are able to be clearly ascertained.  This allows us to help elucidate the physical interpretation of the terms in the quantum statistical mechanics treatment \cite{Floerchinger20}. The time rate of change of the relative quantum entropy is a measure of whether a system is evolving towards or away from the maximally mixed state and the rate at which it does so. Scaling it by the temperature of the state described by $\sigma_m$ gives the time rate of change in the energy associated with non-equilibrium terms.
\end{itemize}


%

\end{document}